\documentclass[twocolumn,aps,pre]{revtex4}

\usepackage{graphicx}
\usepackage{dcolumn}
\usepackage{amsmath}
\usepackage{latexsym}
\usepackage{verbatim}

\begin {document}

\title {Describing Self-organized Criticality as a continuous phase transition}

\author{S. S. Manna}
\affiliation{B-1/16 East Enclave Housing, 02 Biswa Bangla Sarani, New Town, Kolkata 700163, India}

\begin{abstract}
      Can the concept of self-organized criticality, exemplified by models such as the sandpile model, 
   be described within the framework of continuous phase transitions? In this paper, we provide extensive 
   numerical evidence supporting an affirmative answer. Specifically, we explore the BTW and Manna sandpile 
   models as instances of percolation transitions from disordered to ordered phases. To facilitate this 
   analysis, we introduce the concept of `drop density'—a continuously adjustable control variable that 
   quantifies the average number of particles added to a site. By tuning this variable, we observe a 
   transition in the sandpile from a sub-critical to a critical phase. Additionally, we define the scaled size 
   of the largest avalanche occurring from the beginning of the sandpile as the `order parameter' for the 
   self-organized critical transition and analyze its scaling behavior. Furthermore, we calculate the 
   correlation length exponent and note its divergence as the critical point is approached. The finite
   size scaling analysis of the avalanche size distribution works quite well at the critical point of the
   BTW sandpile.
\end{abstract}

\maketitle

\section{Introduction}

      In their seminal paper on Self-Organized Criticality (SOC), Bak, Tang, and Wiesenfeld (BTW) \cite{BTW} 
   proposed that certain natural systems exhibit long-range correlations spontaneously \cite{Guten, Solar, 
   River, Sandpile}. They argued that this phenomenon is the result of a self-organizing process that does 
   not require precise tuning of any control variable. Unlike the laboratory experiments on continuous phase 
   transitions, where fine-tuning is necessary to position the system exactly at the critical point to observe 
   fluctuations across all length and time scales, a SOC system naturally evolves to a stationary state that 
   exhibits critical behavior without such adjustments. This stationary state, characterized by long-range 
   correlations, gives rise to the concept of `self-organized criticality.' The BTW paper illustrated this 
   concept using simple toy models, including the well-known `sandpile model' \cite{BTW, Manna, Dhar1, Dhar2, Manna4}.

      In the study of continuous phase transition, the order parameter plays a crucial role in characterizing the phase 
   of the system \cite {Stanley}. It is a measurable quantity that reflects the degree of order within the system. In the 
   ordered phase, the order parameter has a non-zero value, indicating that the system exhibits some form of long-range 
   order or symmetry breaking. As the system approaches the transition point, the order parameter decreases and eventually 
   becomes zero at the critical temperature (or critical point). In the disordered phase the order parameter remains zero, 
   signifying a lack of long-range order or symmetry breaking. In magnetic systems, the order parameter is typically the 
   total magnetization of the system. In percolation theory, the order parameter is the fractional size of the largest 
   cluster of occupied sites \cite {Stauffer,Saberi,Manna-Ziff}. As the occupation probability $p$ increases and approaches 
   the critical percolation threshold $p_c$, the size of the largest cluster grows from a small fraction of the total number 
   of sites to a significant fraction, marking the transition from a disordered to an ordered state. In the percolation model, 
   the critical occupation probability $p_c$ marks the point at which a phase transition occurs from a state where clusters 
   are too small to span the system to a state where a macroscopic cluster appears.

      In contrast, in the models of SOC, it is not customary to define an order parameter. 
   Here, the appearance of avalanches of all length and time scales is considered to be the signature 
   of the long-range order. The sizes and life-times of avalanches are power-law distributed. That means the 
   small avalanches are too many in number \cite {Manna3} and their number gradually decreases as the avalanche 
   size increases. At the end there are some avalanches of the macroscopic sizes as well. BTW claimed there is no 
   tuning parameter in SOC since the system is attracted to the stationary state following its own dynamical 
   process under the external driving mechanism. Since then, no body talks about a tuning parameter in the 
   SOC system. 

      In spite of such a prevalent scenario we think it may be possible to describe the sandpile model 
   in the framework of the continuous phase transition, specifically the percolation transition and try
   to do a similar formulation in this paper. We first recognize the fact that the external agency that perturbs the 
   system regularly at every tick of the clock may be considered as the generator of the continuously 
   tunable control variable. With this clue, we define a quantity as `drop density' to play the role of 
   control variable. Secondly, borrowing the idea from percolation theory we consider the size of the 
   largest avalanche as the order parameter for the sandpile problem. 

      Some exposure of the previous literature is necessary. The connection between SOC and 
   percolation was first designed in a model called `invasion percolation' introduced four years 
   before SOC was proposed where an interface evolves by invading the weakest site / bond on the 
   interface \cite {Wilkinson}. In the early days of SOC a sandpile had been studied as a continuous
   phase transition \cite {Tang}. In this model a super-critical sandpile was considered whose slope 
   is larger than the critical slope and this case is different from what we consider here. Some 
   more study and discussions on the relation between continuous phase transitions and self-organized 
   criticality are available in \cite {Tang1,Lubeck}. 

      Secondly, in a conservative stochastic sandpile grown within a closed boundary system, there is 
   a continuous phase transition from the `absorbed' to `active' phase first introduced and studied
   by Vespignani et. al. \cite {Dickman1,Dickman2}. In such a closed system, the total 
   number of particles is a constant parameter of the dynamics. Instead of the sand grains they 
   interpreted them as the energy and therefore since it is a conservative system, the overall energy 
   density $\zeta = E/L^d$ is a constant of motion. They also considered the continuum description of the
   sandpile model and defined the local space and time dependent energy density $\zeta({\bf \hat x},t)$. 
   One therefore considers infinite number of different closed systems for infinite number of different 
   values of the parameter $\zeta$ and then considers $\zeta$ as the control variable. For the sandpile 
   interpretation of the conservative system a quantity $\rho$, the average number of sand particles per 
   site, is the control variable which has the latest updated value of the critical density $\rho_c 
   \approx$ 2.125288 \cite {Fey}. Therefore, our control variable `drop density' $\tau(t)$ is similar to 
   their energy density $\zeta$ but which has not been defined till now for the open sandpile which enables 
   its description as a continuous transition. 

      A detailed look into the interpretation of SOC and its relation
   difference with a slowly driven, interaction dominated threshold (SDIDT) had been studied in 
   \cite {Watkins}. More recently, over the last decade the phenomenon of `Hyperuniformity' has 
   been discovered in the Manna class of sandpiles. It primarily deals with the dependence of the 
   fluctuations in the particle density within a small box of the sandpile on the lateral size of 
   the box \cite {Hexner,Tjhung,Wilken,Punya1,Punya2}.

\begin{figure}[t]
\includegraphics[width=6.0cm]{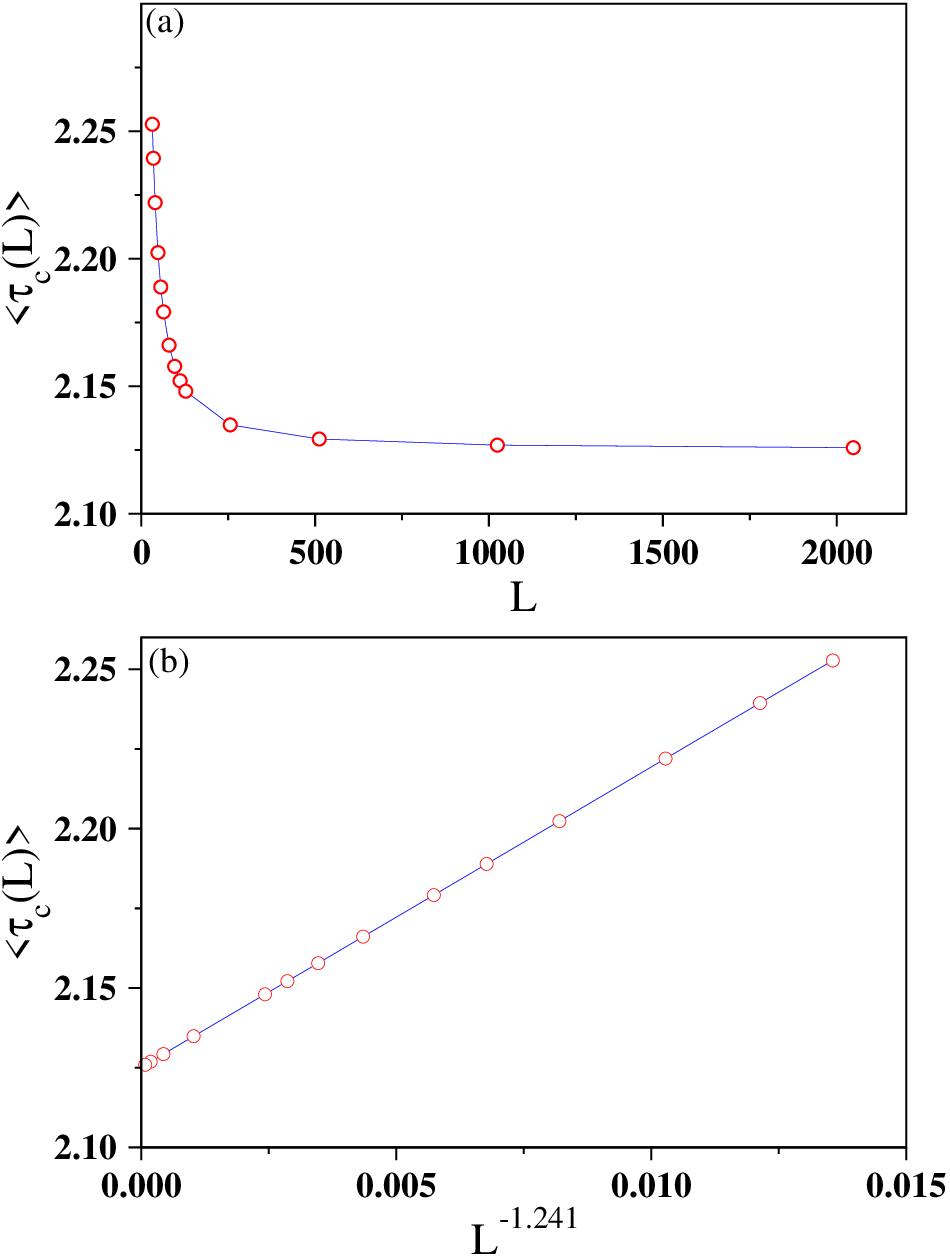}
\caption{BTW sandpile:
   (a) The average value of critical drop density $\langle \tau_c(L) \rangle$ estimated using the first time 
   spanning avalanches have been plotted against the system size $L$.
   (b) The same values of $\langle \tau_c(L) \rangle$ have been plotted against $L^{-1.241}$ to obtain the
	best fitted straight line $\langle \tau_c(L) \rangle = 2.1252 + 9.396.L^{-1.241}$
	so that $\tau_c$ = 2.1252 and $1/\nu = 1.241$.
}
\label{FIG01}
\end{figure}

      Now we start with a brief description of the BTW sandpile. Sandpiles are the prototypical 
   models of SOC \cite {BTW,Manna,Dhar,Frontiers,Stella2,Chessa1,Pietro1,Lubeck1,Biham1,Manna2,
   Dickman,Garber}. Random dropping of sand particles on to the lattice creates density 
   inhomogeneities at different locations. The dynamical evolution process of the sandpile has an 
   inbuilt diffusion mechanism so that the sand particles get distributed within the lattice to 
   smooth out the coarse grained landscape. Rules for the BTW sandpile are as follows: On an empty 
   square lattice of size $L \times L$ sand particles are dropped only one at a time. Multiple 
   particles are allowed to share the same lattice site at the same time, but only up to a certain 
   extent. Let the number $h(i,j)$ of particles denote the height of the sand column at the lattice 
   site $(i,j)$. The addition of a sand particle therefore implies:
\begin {center}
\verb !h(i,j)! $\rightarrow$ \verb !h(i,j) + 1.!
\label {EQN01}
\end {center}
      A sand column is said to be unstable only when its height is greater than a threshold value $h_s-1$, 
   e.g., $h_s$ = 4 for the square lattice. An unstable sand column topples and transfers $4$ particles, 
   one each to the neighbouring sites:
\begin {center}
\begin {tabular}{c}
\verb !h(i,j)! $\rightarrow$ \verb !h(i,j) - 4! \\
\verb !h(i!$\pm1$, \verb!j!$\pm1$) $\rightarrow$ \verb!h(i!$\pm1$, \verb!j!$\pm1$) + 1.
\end {tabular}
\label {EQN02}
\end {center}
   After receiving the sand particles, one or more neighbouring sites may also topple in a similar fashion. 
   In this way a cascade of toppling events follows in succession, called an 
   `avalanche'. A detailed description of the heights of the sand columns at all sites of the lattice constitute a 
   microstate of the sandpile. Such a microstate is referred as stable, if none of the sand columns is unstable, 
   otherwise it is said to be unstable. An avalanche takes the system from one stable state to another. There 
   are a number of popular measures for the avalanche sizes. Most popularly, the avalanche size is measured by the total 
   number $s$ of sand column topplings. Another measure of the avalanche size is 
   the total number $d$ of distinct sites toppled at least once during the avalanche.

\begin{figure}[t]
\includegraphics[width=6.0cm]{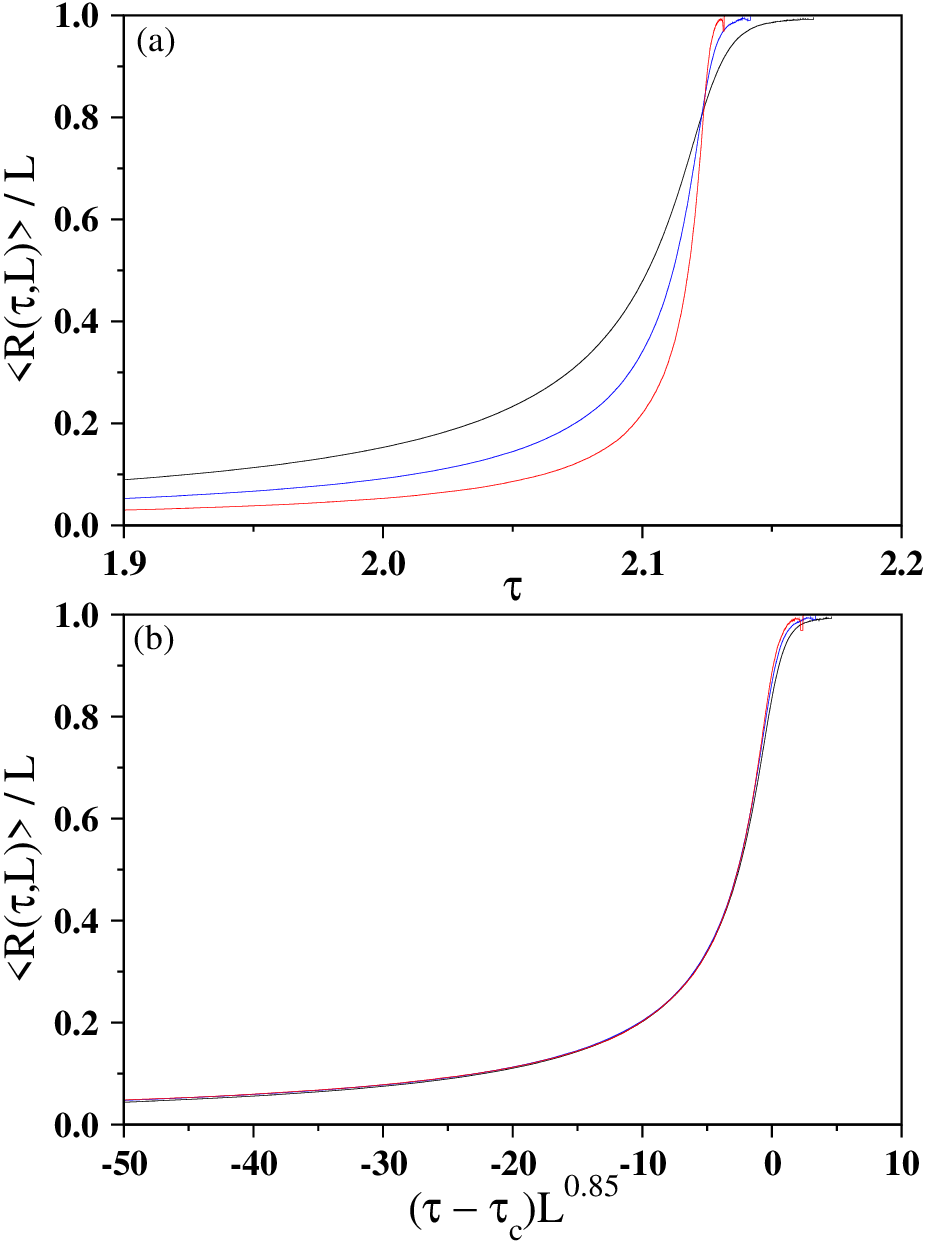}
\caption{BTW sandpile:
   (a) The average value $\langle R(\tau,L) \rangle$ of the maximal physical sizes of the avalanches
   scaled by the lattice size $L$ have been plotted against the drop density $\tau$ for three 
   different lattice sizes.
   (b) A scaling plot of the same data have exhibited a data collapse for all three system sizes
   using $\tau_c = 17/8$.
}
\label{FIG02}
\end{figure}
\begin{figure}[t]
\includegraphics[width=6.0cm]{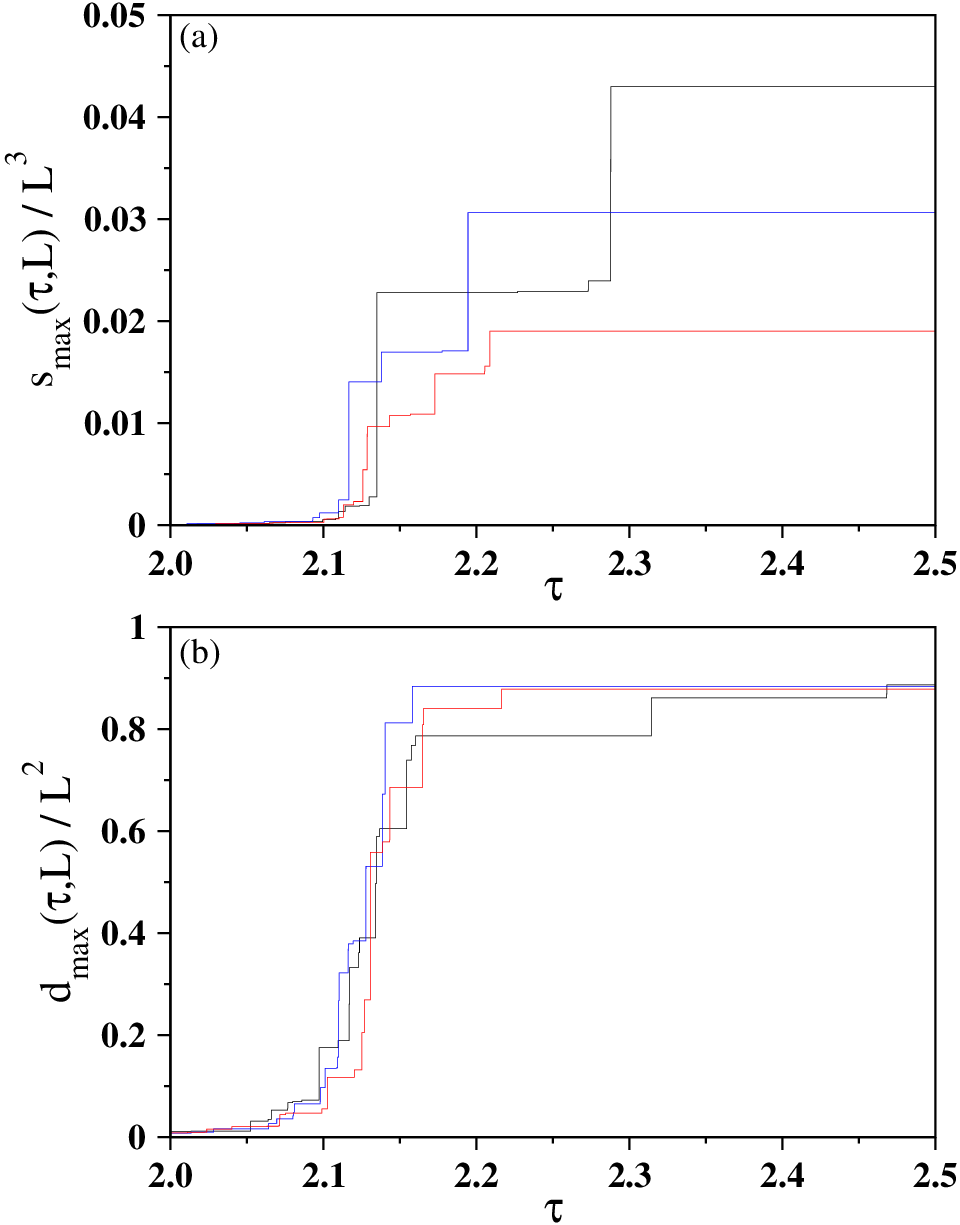}
\caption{
   BTW sandpile on the square lattice of size $L$ = 256.
   (a) Plot of the scaled value $s_{max}(\tau,L) / L^3$ of the largest avalanche size and
   (b) plot of the scaled value $d_{max}(\tau,L) / L^2$ of the largest avalanche size
   measured by the total number of distinct sites, against the drop density $\tau$. Three
   curves are for three independent sandpiles evolving from the empty lattice.
}
\label{FIG03}
\end{figure}
\begin{figure}[t]
\includegraphics[width=6.0cm]{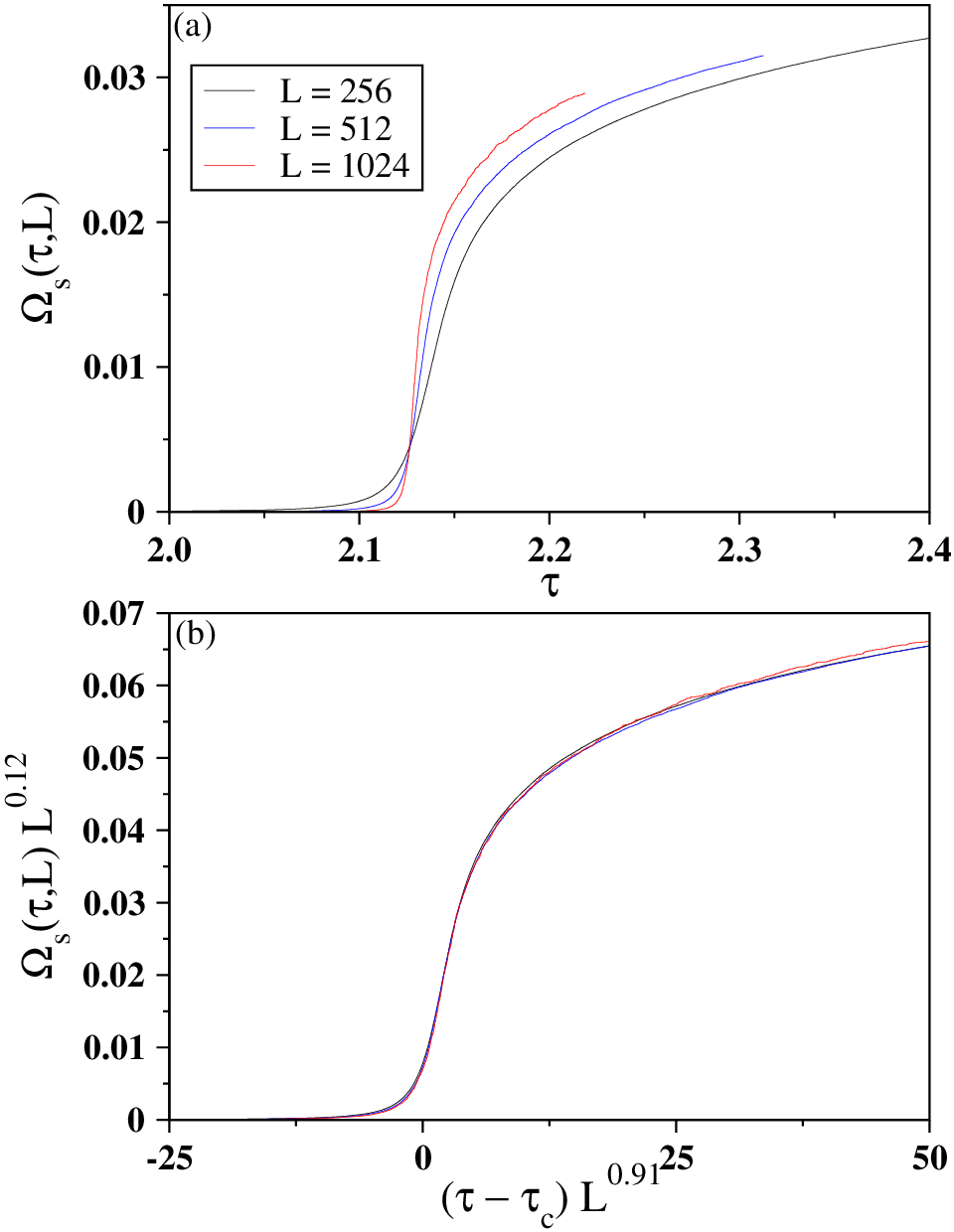}
\caption{BTW sandpile:
   (a) The order parameter $\Omega_s(\tau,L)$ defined by Eqn. \ref {FIG03} has been plotted against 
   the drop density $\tau$ for sandpiles on three different lattice sizes.
   (b) The same data have been scaled and plotted as: $\Omega_s(\tau,L) L^{0.12}$ vs. $(\tau - \tau_c) L^{0.91}$
   using $\tau_c = 2.125$ to obtain a nice collapse of the plots on top of one another.
}
\label{FIG04}
\end{figure}
\begin{figure}[t]
\includegraphics[width=6.0cm]{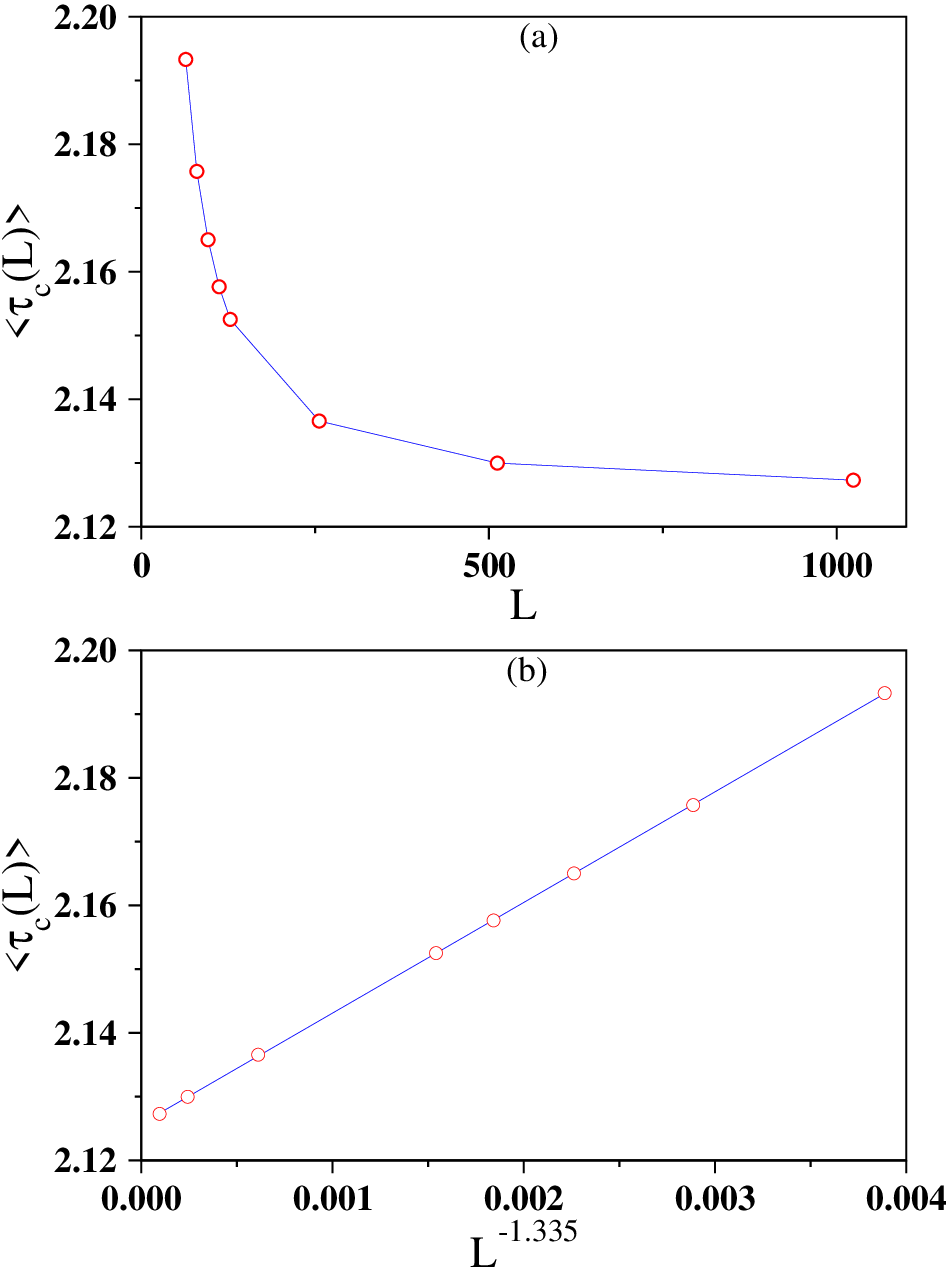}
\caption{BTW sandpile:
   (a) The average values of the critical drop density $\langle \tau_c(L) \rangle$ corresponding
       to the largest jumps in the order parameter $\Omega_d(d,L)$ have been plotted against the 
       lattice size $L$.
   (b) The same values of $\langle \tau_c(L) \rangle$ have been plotted against $L^{-1.335}$ to
       obtain the best fitted straight line which extrapolates to $\tau_c = 2.1258$ and $1/\nu$= 1.335.
}
\label{FIG05}
\end{figure}
\begin{figure}[t]
\includegraphics[width=6.0cm]{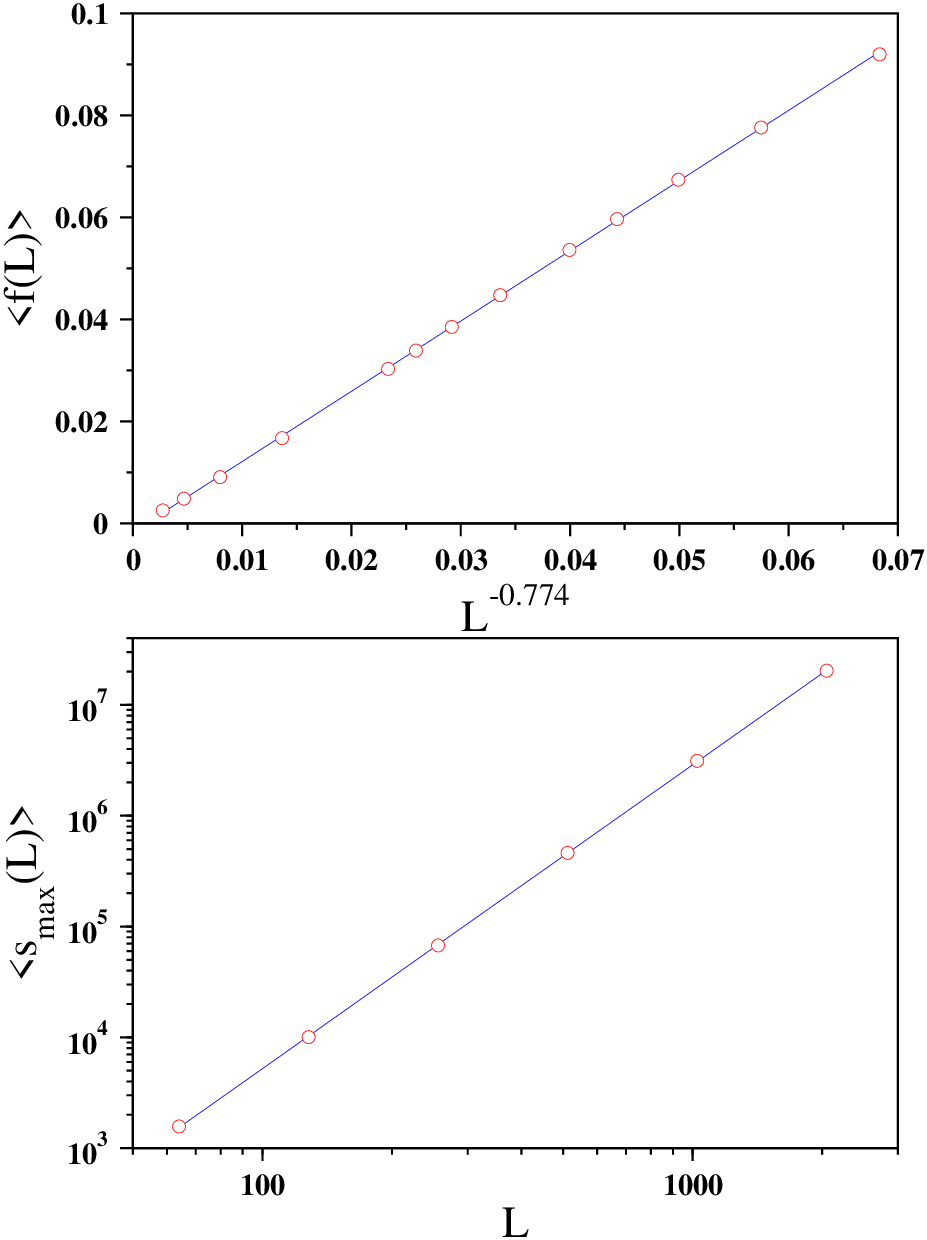}
\caption{BTW sandpile:
   Two quantities have been estimated when the drop density is $\tau = 17/8$.
   (a) The net outflux per site is best fitted to a form $\langle f(L) \rangle = f_c + A L^{-0.774}$ with 
       $f_c$ = -0.0016.
   (b) The average size of the largest avalanche grows as $\langle s_{max}(L) \rangle \sim L^{2.741}$.
}
\label{FIG06}
\end{figure}

      The boundaries of the system are open and the sand columns which topple at the boundary transfer one 
   or two particles outside the boundary. This way the system looses sand mass and it constitutes 
   the outflow current. Since we start from the empty lattice, initially most of the dropped particles are absorbed directly, since 
   the average height of the sand columns at this stage is very small, and hardly any sand column 
   reaches the threshold height. Therefore, at the early stage, the typical avalanche sizes are 
   small. However, gradually the average height increases and the avalanche sizes grow systematically.

\begin{figure}[t]
\includegraphics[width=6.0cm]{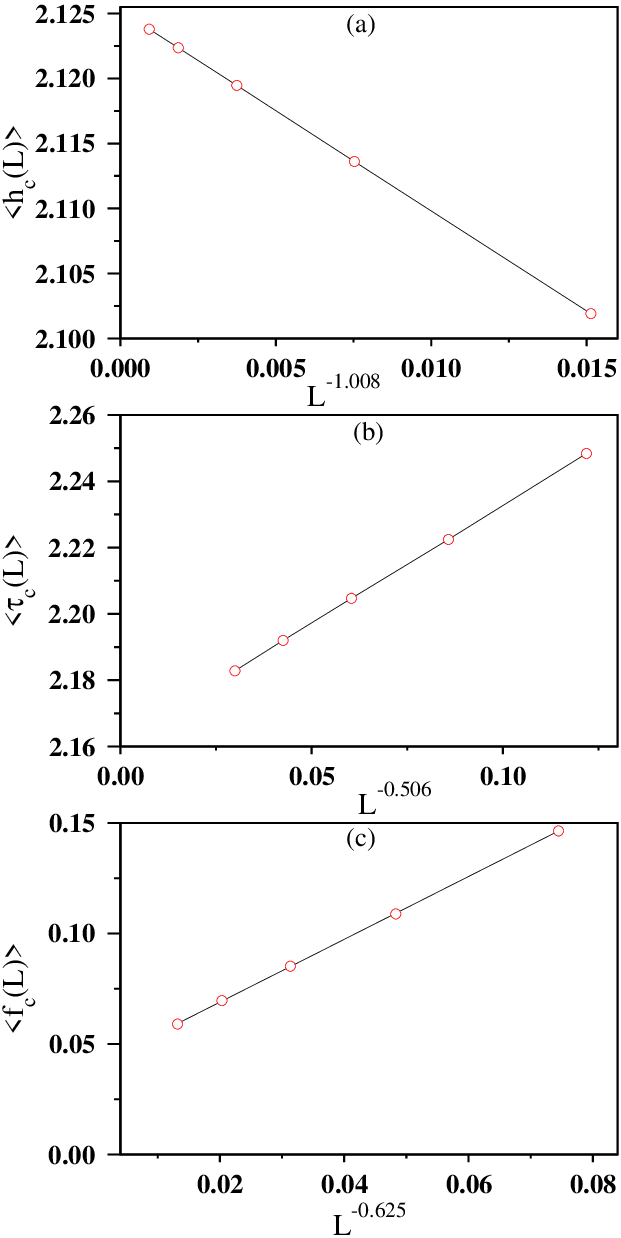}
\caption{BTW sandpile:
   Using the burning algorithm we detect the first appearance of the recurrent stable state when we 
   terminate the evolution of the sandpile. The corresponding drop density $\tau_c(L)$ is the 
   critical drop density for the run. Using this method, the average values of three quantities have
   been estimated over many such independent runs:
   (a) the average critical height $\langle h_c(L) \rangle$ of the sand column,
   (b) the average critical drop density $\langle \tau_c(L) \rangle$, and
   (c) the average outflux $\langle f_c(L) \rangle$. On finite size extrapolations, the asymptotic 
   values are found to be $h_c$ = 2.1252, $\tau_c$ = 2.1616, $f_c$ = 0.0405 respectively.
}
\label{FIG07}
\end{figure}

      The dynamical evolution of the sandpile is followed by adding one particle at a time. 
   Irrespective of the size of the avalanche so generated, the system is then allowed to relax 
   completely till all the activities die out and a new stable state is reached, and only then the next particle is dropped. It is assumed 
   that all avalanches propagate infinitely fast and take only zero time. This implies that sand
   additions create a constant inflow current of strength unity. One defines a `pseudo time' $t$ 
   which is equal to the total number of sand particles dropped onto the lattice. Eventually, after 
   some characteristic relaxation time $t_c(L) \sim L^2$ the average magnitude of the 
   outflow current also reaches unity when the inflow and the outflow currents become 
   equal and at that time the system arrives at the stationary state. One defines an  
   instantaneous average height $h_{av}(t,L)$ over all sites of the lattice. During 
   the simulation of a specific sandpile starting from an empty lattice sand particles 
   are dropped one at a time at the randomly selected locations and when this process 
   is followed ad infinitum, we refer it as a `run'. Therefore, the average height of
   a sand column over many runs is denoted by $\langle h(t,L) \rangle$ which attains
   the time independent value $\langle h(L) \rangle$ in the stationary state. The 
   instantaneous average height fluctuates around the steady average value. It has been 
   shown analytically that the steady average height for the infinitely large square 
   lattice is $h_c = \lim_{L \to \infty} \langle h(L) \rangle = 17/8$ \cite {Jeng,Caracciolo}.

      A stable microstate of the BTW sandpile can be of two types, namely, the `recurrent' 
   and the `transient' \cite {Dhar}. Only the recurrent states, and not the transient states,
   appear with uniform probability in the stationary state. For example, an empty lattice 
   is a transient state of the sandpile. When a sandpile is evolved 
   from such a state, it passes through a sequence of transient states. After the
   characteristic relaxation time $t_c(L)$ the sandpile eventually lands up at a recurrent 
   state, and then on all stable states are recurrent states only. These recurrent states 
   constitute the stationary state of the sandpile. In the asymptotic limit of $L \to \infty$ this
   stationary state is recognized as the critical 
   state since fluctuations of all length and time scales appear in this
   state. Question is, how to identify if a stable state is recurrent or transient? Fortunately, 
   there is a precise algorithm to answer this question \cite {Satya}, called the `burning' 
   algorithm. Using this algorithm one can identify a completely burnable stable state as
   the recurrent state.

      In section II we describe the BTW sandpile dynamics as the continuous phase transition. In section 
   III we present a similar description for the Manna sandpile as well. In section IV we summarize.

\section {BTW Sandpile}

      We ask at this point, can one define a control variable which may be tuned continuously to describe 
   the arrival of the sandpile to the self-organized critical state, starting from the sub-critical state of a completely empty lattice? 
   Here we have introduced such a variable 
\begin {equation}
\tau(t) = t / L^2
\label {EQN03}
\end {equation}
   and refer it as the `drop density', since it is 
   the average number of sand particles dropped at a typical site of the lattice. Unlike the average height 
   of the sand column which grows to reach its stationary state value $\langle h(L) \rangle$, the value of 
   $\tau(t)$ grows forever after. 

\begin{figure}[t]
\includegraphics[width=6.0cm]{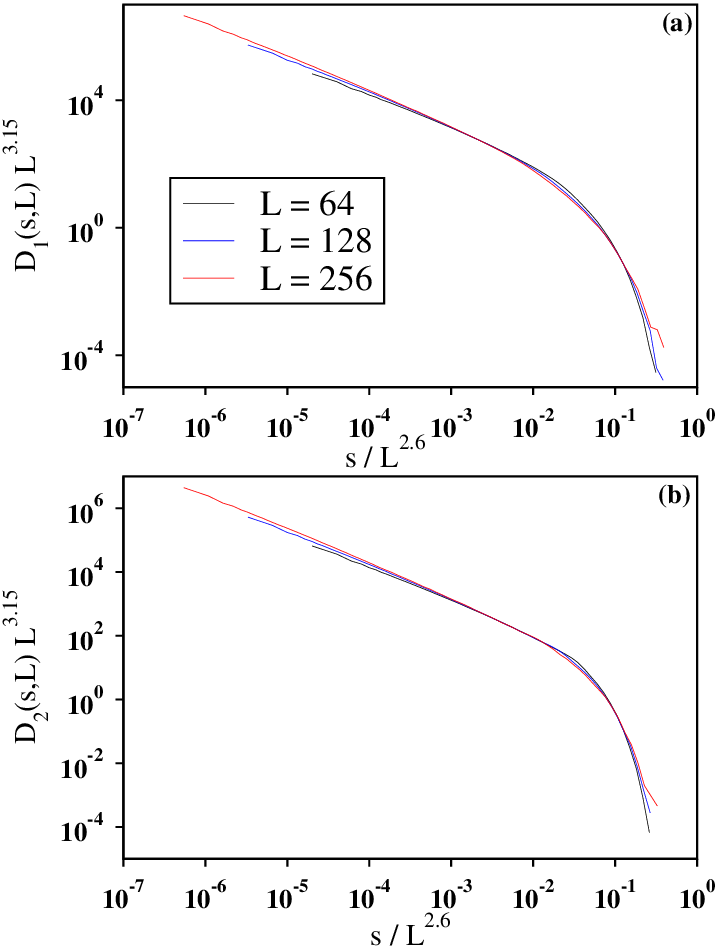}
\caption{BTW sandpile: 
   The finite size scaling analyses of the avalanche size distributions using method:
   (a) the size of the avalanche exactly at time $t = \tau_cL^2$, and
   (b) the size of the avalanche right before the first spanning avalanche.
}
\label{FIG08}
\end{figure}

\begin{figure}[t]
\includegraphics[width=6.0cm]{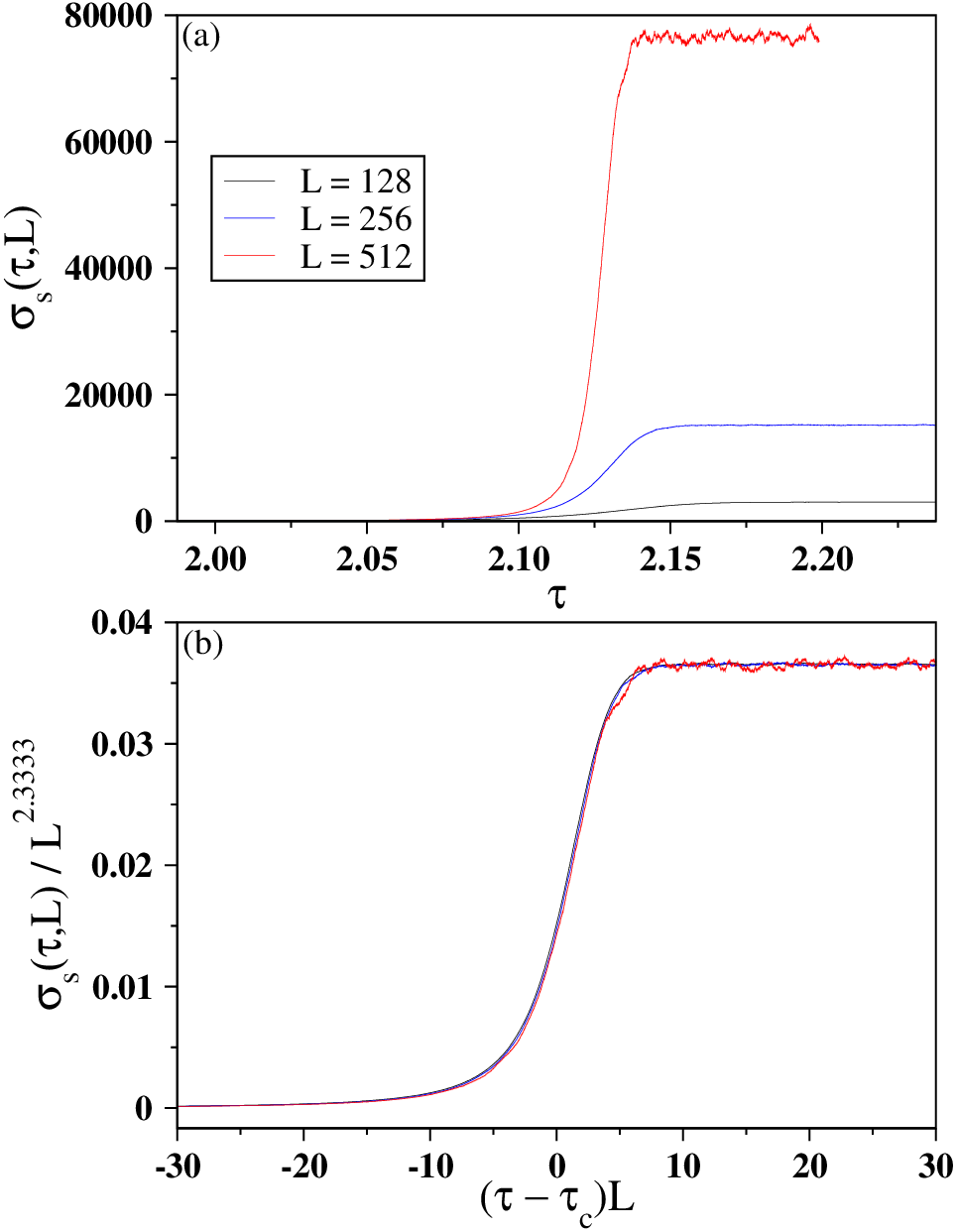}
\caption{BTW sandpile: 
   (a) The standard deviation $\sigma_s(\tau,L)$ of avalanche sizes created for the each value of 
   $\tau$ but for large number of independent runs have been plotted against the $\langle h(L) \rangle$
   drop density $\tau$ for three lattice sizes. 
   (b) The same data have been scaled and re-plotted as: $\sigma_s(\tau,L) / L^{2.3333}$ vs. $(\tau - \tau_c)L$
   with $\tau_c$ = 2.125 to obtain a nice collapse of the data.
}
\label{FIG09}
\end{figure}

      The critical value of the drop density $\tau_c(L)$ is defined to be the value of $\tau$ when 
   the system moves into the stationary state. We have defined $\tau_c(L)$ for a single run in three different 
   ways. First, we define the `spanning avalanches' analogous to the `spanning clusters' in percolation 
   theory. This way we can estimate asymptotic values of both the drop density $\tau_c$ and the correlation 
   length exponent $\nu$. A spanning avalanche is the one which touches at least one of the two opposite 
   boundaries. Once we have such an avalanche for the first time, we note down the value of the drop density 
   $\tau_c(L)$ and terminate the simulation. The average critical drop density $\langle \tau_c(L) \rangle$ 
   have been determined over many runs and for different lattice sizes $L$ starting from 32 to 2048, $L$ 
   being increased by a factor of 2.

      In Fig. \ref {FIG01}(a) we have plotted $\langle \tau_c(L) \rangle$ against $L$ and observe 
   that as the system size increases the average value of the critical drop density gradually decreases
   and slowly approaches to its asymptotic value $\tau_c$. Secondly, in Fig. \ref {FIG01}(b) we have plotted 
   the same values of $\langle \tau_c(L) \rangle$ against $L^{-1.241}$ to obtain the best fitted 
   straight line:
\begin {equation}
\langle \tau_c(L) \rangle = \tau_c + AL^{-1/\nu}
\label {EQN04}
\end {equation}
   where $\tau_c = 2.1252$, $1/\nu$ = 1.241 and A = 9.396. We compare the asymptotic value of $\tau_c$
   with the exact value of the critical height $h_c = 17/8$ \cite {Jeng} for the BTW model on square 
   lattice and the difference is 0.0002. Therefore, we conjecture that $\tau_c = h_c = 17/8$ and in the 
   following we report that other estimates also support this conjecture. Similarly, the value of the
   reciprocal of the correlation length exponent is found to be $1/\nu$ = 1.241 which leads us to conjecture
   $\nu = 4/5$. This value may be compared with $\nu = 4/3$ \cite {Stauffer} for the ordinary percolation 
   in two dimension.

      The physical size of an avalanche is defined to be the larger of its extensions along the $x$ and the 
   $y$ axes. After the addition of every sand particle the physical size of the corresponding avalanche is 
   measured. We have calculated the running value $R(\tau,L)$ of the maximal physical size of all avalanches
   starting from the beginning of the sandpile till the drop density $\tau$. In Fig. \ref {FIG02}(a) we have 
   plotted the scaled average maximal physical size 
   $\langle R(\tau,L) \rangle / L$ against $\tau$ for three system sizes $L$ = 256 (black), 512 (blue), and 
   1024 (red). Successive curves meet at points whose $x$ coordinates approach $\tau_c$ in the limit of $L \to \infty$.
   This prompted us to try a finite size scaling analysis of the same data in Fig. \ref {FIG02}(b). Using the asymptotic 
   critical value $\tau_c = 17/8$ we have scaled the $x$-axis as $(\tau - \tau_c)L^{0.85}$ and the $y$-axis 
   as $\langle R(\tau,L) \rangle / L$. The data collapse is quite good. The scaling form is therefore,
\begin {equation}
\langle R(\tau,L) \rangle / L \sim {\cal G}[(\tau - \tau_c)L^{0.85}],
\label {EQN05}
\end {equation}
   here ${\cal G}(x)$ is an universal scaling function which should diverge as $x^{-\delta}$ in the $x \to 0$ limit
   i.e., in the asymptotic limit of $L \to \infty$. In this limit $\langle R(\tau) \rangle \sim (\tau - \tau_c)^{-\delta}$
   implies $1 - 0.85\delta = 0$, or $\delta \approx$ 1.176.

      In the sandpile model the appearance of large avalanches is an indication of high level 
   of order being set in the system. Starting from the initial empty lattice the sizes of the 
   avalanches gradually grow as the drop density increases. A large avalanche covering a large
   area of the system indicates the appearance of the long range correlation. Similar to the 
   theory of percolation phenomena we have defined the order parameter:
\begin {equation}
\Omega_s(\tau,L) = \langle s_{max}(\tau,L) \rangle / L^3.
\label {EQN06}
\end {equation}
   Here, $s_{max}(\tau,L)$ is the maximal of the sizes of all avalanches generated between 
   time $t$ = 1 to $t = \tau L^2$. The maximal cluster size has been scaled by $L^3$ since the
   maximum size of all possible avalanches for a sandpile grows as $L^3$ \cite {Garber}. A scaled 
   value $s_{max}(\tau,L) / L^3$ against $\tau$ has been plotted for three independent runs in Fig. 
   \ref {FIG03}(a) for $L$ = 256. Its value is very small till the drop density $\tau \sim \tau_c(L)$. 
   However, beyond this point it increases very fast in large steps. Since it is the running value 
   of the largest avalanche size, it is a monotonically increasing function of $\tau$. Another point
   we must mention that in ordinary continuous transition the order parameter assumes a zero value
   at the critical point where long range fluctuations are observed. In comparison, in SOC or say in
   the sandpile models, the entire stationary state is critical where fluctuations of all length
   scales are observed. In that comparison, our order parameter as defined above is perhaps 
   a quantity that is very ``similar'' to order parameters defined for continuous phase transitions.

      The order parameter $\Omega_s(\tau,L)$ has been plotted in Fig. \ref {FIG04}(a) for different 
   lattice sizes. The three curves for $L$ = 256, 512, and 1024 separate out beyond $\tau_c$. We have 
   therefore scaled both the axes and re-plotted in Fig. \ref {FIG04}(b). First the $x$-axis has been
   transformed as $\tau - \tau_c$ with $\tau_c$ = 17/8. Next the $x$-axis has been scaled by $L^{0.91}$
   and the $y$-axis has been scaled by $L^{0.12}$. A good data collapse have been observed as:
\begin {equation}
\Omega_s(\tau,L) L^{0.12} \sim {\cal F}[(\tau - \tau_c)L^{0.91}].
\label {EQN07}
\end {equation}
   In the asymptotic limit of $L \to \infty$ the order parameter takes the form $\Omega_s(\tau) \sim (\tau - \tau_c)^{\beta}$.
   The Eqn. \ref {EQN07} yields this $L$ independent form if and only if the universal function ${\cal F}(x) \to x^{\beta}$
   so that the power of $L$ is zero, which implies $-0.12 + 0.91 \beta = 0$ or $\beta = 0.132$. This value of
   $\beta$ is very close to the order parameter exponent $\beta = 5/36 \approx 0.139$ of two dimensional percolation 
   problem \cite {Stauffer}.

      Moreover, the order parameter can also be defined using the size $d$ of the avalanches:
   $\Omega_d(\tau,L) = \langle d_{max}(\tau,L) \rangle / L^2$. The advantage here is, like percolation, 
   this order parameter can asymptotically approach to its maximum value unity only. As $\tau$ is 
   gradually increased, the order parameter has to undertake many jumps. Therefore, analogous to the 
   percolation problem we define the value of the critical drop density $\tau_c(L)$ as specific drop density
   when the order parameter gets its maximal jump. An average value 
   $\langle \tau_c(L) \rangle$ of the quantity has been estimated for many independent runs and 
   has been extrapolated as $\langle \tau_c(L) \rangle = \tau_c + AL^{-1/\nu}$. In Fig. \ref {FIG05}(a)
   we have plotted $\langle \tau_c(L) \rangle$ against $L$ for $L$ = 64 to 1024. The same data have been
   re-plotted in Fig. \ref {FIG05}(b) against $L^{-1/\nu}$ and the best fit straight line yields values
   $\tau_c = 2.1258$ and $1/\nu = 1.335$.

      We have also estimated two related quantities when the drop density $\tau$ reaches the critical point
   $\tau_c = 17/8$. The first is the net amount $f(L)$ of outflux of sand mass per site which is the total 
   number of particles that left the system through the boundaries divided by the lattice size $L^2$. In Fig. 
   \ref {FIG06}(a) we have shown a plot of $\langle f(L) \rangle$ against $L^{-0.774}$ and the best fitted 
   straight line has the form $\langle f(L) \rangle = f_c + AL^{-0.774}$ with $f_c = -0.0016$ which is
   very small and close to zero. It implies that only a vanishing fraction of particles leave the system when 
   drop density $\tau$ reaches 17/8. Therefore, the asymptotic value of the critical height in the critical stationary 
   state, i.e., $\tau_c = h_c$. This equality has been numerically verified by the three estimates of $\tau_c$ 
   of the BTW sandpile and two estimates of the Manna sandpile as described in the following. This calculation 
   makes it apparent that for the sandpile model of SOC, the presence of a fixed boundary is rather redundant,
   which has also been observed in \cite {Manna5}. Secondly, we have also calculated the average value of the 
   largest avalanche size $\langle s_{max}(L) \rangle$ till $\tau$ = 17/8 and plotted against $L$ using the 
   double logarithmic scale in Fig. \ref {FIG06}(b) for the range $L$ = 64 to 2048. The plot fits nicely to a 
   straight line with average slope 2.741 i.e., $\langle s_{max}(L) \rangle \sim L^{2.741}$. This exponent 
   is analogous to the fractal dimension $d_f = 91/48$ of the percolation clusters in two dimension at the 
   percolation threshold.

\begin{figure}[t]
\includegraphics[width=6.0cm]{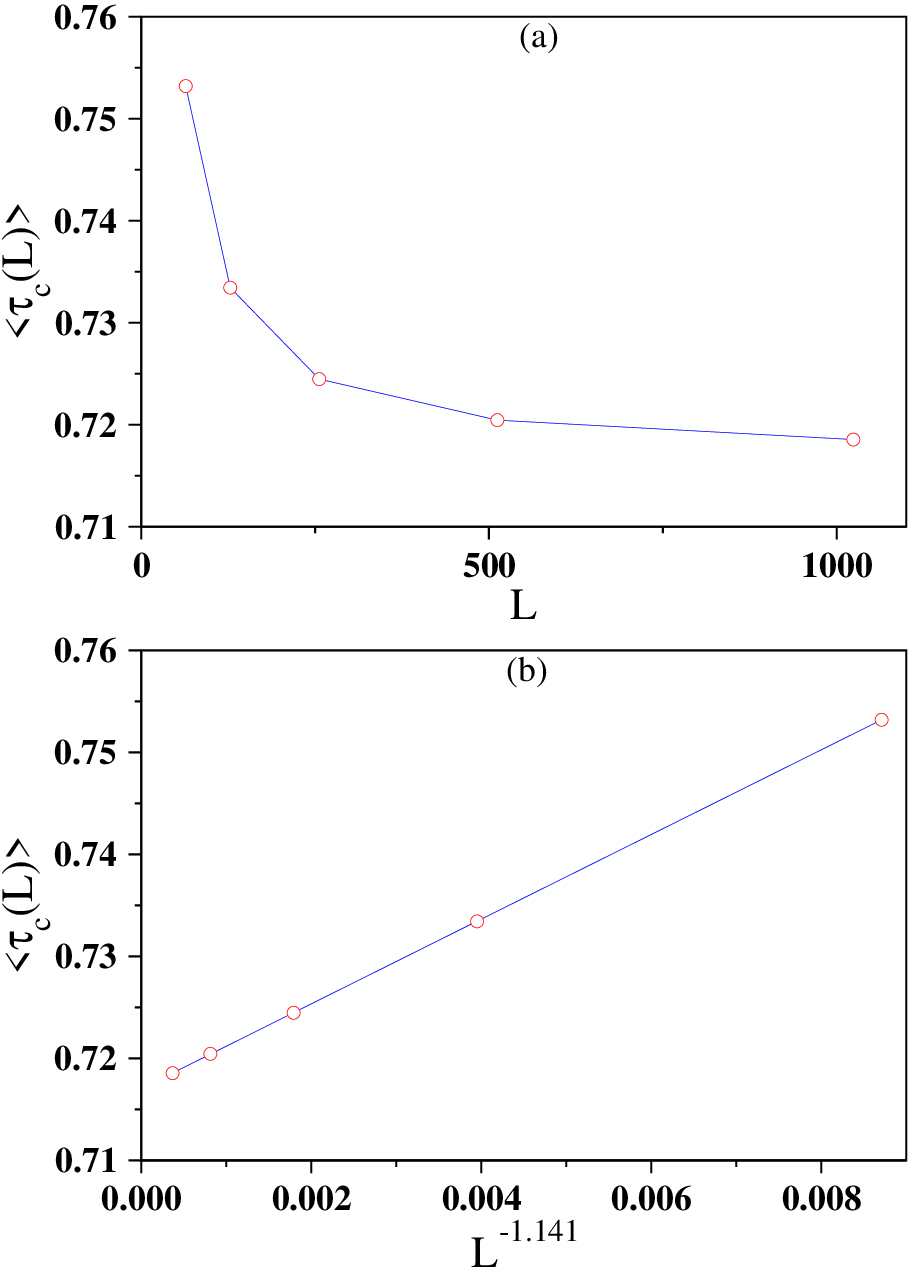}
\caption{Manna sandpile:
   (a) The average value of the critical drop density $\langle \tau_c(L) \rangle$ has been
       estimated using the spanning avalanches.
   (b) The same values of $\langle \tau_c(L) \rangle$ have been plotted against $L^{-1.141}$ to
       obtain the best fitted straight line which extrapolates to $\tau_c = 0.7170$ and $1/\nu$= 1.141.
}
\label{FIG10}
\end{figure}

      We have also used a third definition to calculate the critical drop density $\tau_c$. It is the 
   minimal value of $\tau$ whose corresponding stable state is recurrent. This recurrent state 
   has been identified using the burning algorithm. Starting from an empty square lattice the sandpile has been 
   grown by adding particles one at a time on to the lattice. For each particle added, the avalanche 
   propagates in the medium until it dies completely when the next stable state appears. At that moment 
   a burning process is initiated from outside the boundary. Starting from the boundary the fire front 
   recursively burns the lattice sites. The first time when all sites of the lattice are burnt, the 
   system is considered to arrive at a recurrent state or the stationary state. For every run we have
   stopped at this point and estimated the average height $h(L)$ per site, the drop density $\tau(L)$, 
   and the outflux i.e., the total number of particles dropped out of the system per lattice site till this moment. 
   These quantities have been averaged over a large ensemble of independent runs to obtain the average 
   values $\langle h(L) \rangle$, $\langle \tau(L) \rangle$ and the average of the net outflow 
   $\langle f(L) \rangle$.

      The entire simulation has been repeated for a number of lattice sizes, from $L$ = 64 to 1024. In 
   Fig. \ref {FIG07} we have shown the finite size extrapolations of three quantities to their asymptotic limits: 
   (a) plot of $\langle h_c(L) \rangle$ against $L^{-1.01}$ extrapolates to $h_c = 2.1252$,
   (b) plot of $\langle \tau_c(L) \rangle$ against $L^{-0.506}$ extrapolates to $\tau_c = 2.1616$, and
   (c) plot of $\langle f_c(L) \rangle$ against $L^{-0.625}$ extrapolates to $f_c = 0.0405$.
   We understand that the matching of $\tau_c$ with the actual value $h_c = 17/8$ has not been that good as
   in previous two estimates, and also the value of $f_c$ is not that close to zero which we believe may be 
   due to finite size of the system studied in this analysis.

\begin{figure}[t]
\includegraphics[width=6.0cm]{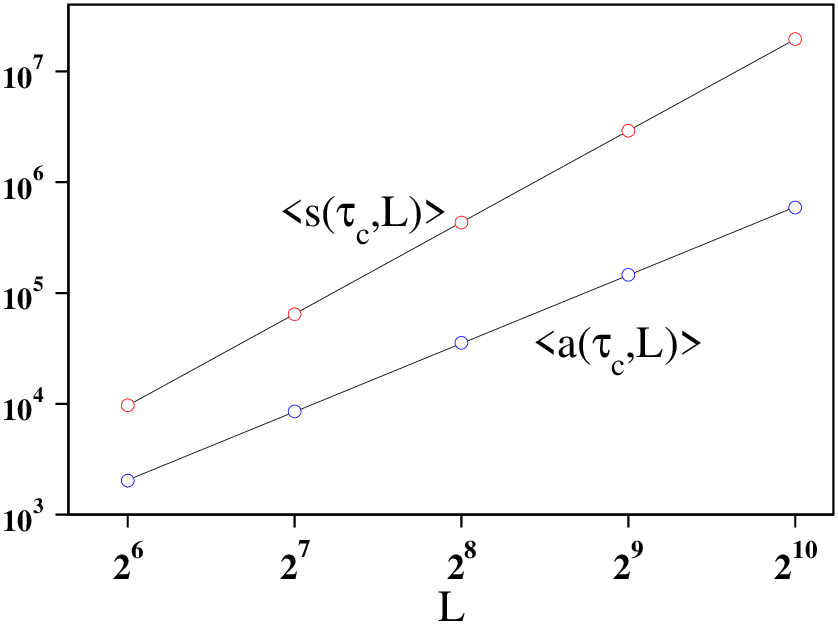}
\caption{Manna sandpile: 
      The average size of the first spanning avalanches $\langle s(\tau_c,L) \rangle$ and
   the average area of the spanning avalanches $\langle a(\tau_c,L) \rangle$ have been plotted 
   against the lattice size $L$ using the $\log-\log$ scale. Linear fits yield 
   $\langle s(\tau_c,L) \rangle \sim L^{2.745}$ and $\langle a(\tau_c,L) \rangle \sim L^{2.047}$.
}
\label{FIG11}
\end{figure}

      Finally, we have estimated the probability distribution of avalanche sizes using two different 
   methods. In the first method we have terminated every run right after the drop density $\tau = 
   \tau_c = 17/8$ has been reached and collect the data for the size of the last avalanche. The 
   distribution of these avalanche sizes is denoted by $D_1(s,L)$. Since only one size data for the 
   entire run is collected, we had to simulate a large number of runs, namely, 99.95 millions for 
   $L$ = 64 to 4.07 million for $L$ = 256. In Fig. \ref {FIG08}(a) we have plotted $D_1(s,L) L^{\zeta}$ 
   against $s / L^{\eta}$ with $\zeta=3.15$ and $\eta=2.60$ to obtain a good collapse of the data 
   confirming the scaling form
\begin {equation}
D_1(s,L) L^{\zeta} \sim {\cal H}[s / L^{\eta}].
\end {equation}
   Since it is assumed that asymptotically, the avalanche size distribution is a simple power law, 
   like: $D_1(s) \sim s^{-\kappa_1}$, we get $\kappa_1 = \zeta / \eta = 3.15 / 2.6 \approx 1.21$. 
   In the second method the simulation is stopped at different values of $\tau$ for different runs. 
   A typical run is terminated right after a spanning avalanche appears for the first time and we have 
   collected the size data for just the previous avalanche. We assume for the second size as:
   $D_2(s) \sim s^{-\kappa_2}$ and again from the finite size scaling analysis in Fig. \ref {FIG08}(b) 
   we estimate $\kappa_2 \approx 1.21$. We conclude that the avalanche size exponent $\kappa 
   \approx 1.21$ which is consistent with previous estimates of 1.22 and 1.20 for the avalanche 
   size exponent \cite {Manna3}. Here, the finite size scaling analysis of the distributions worked
   quite well. One point must be noted that for BTW model the avalanche size distribution determined 
   in the stationary state had been notorious for not obeying the scaling analysis. It seemed this 
   is not so for the present method of analysis.

\begin{figure}[t]
\includegraphics[width=6.0cm]{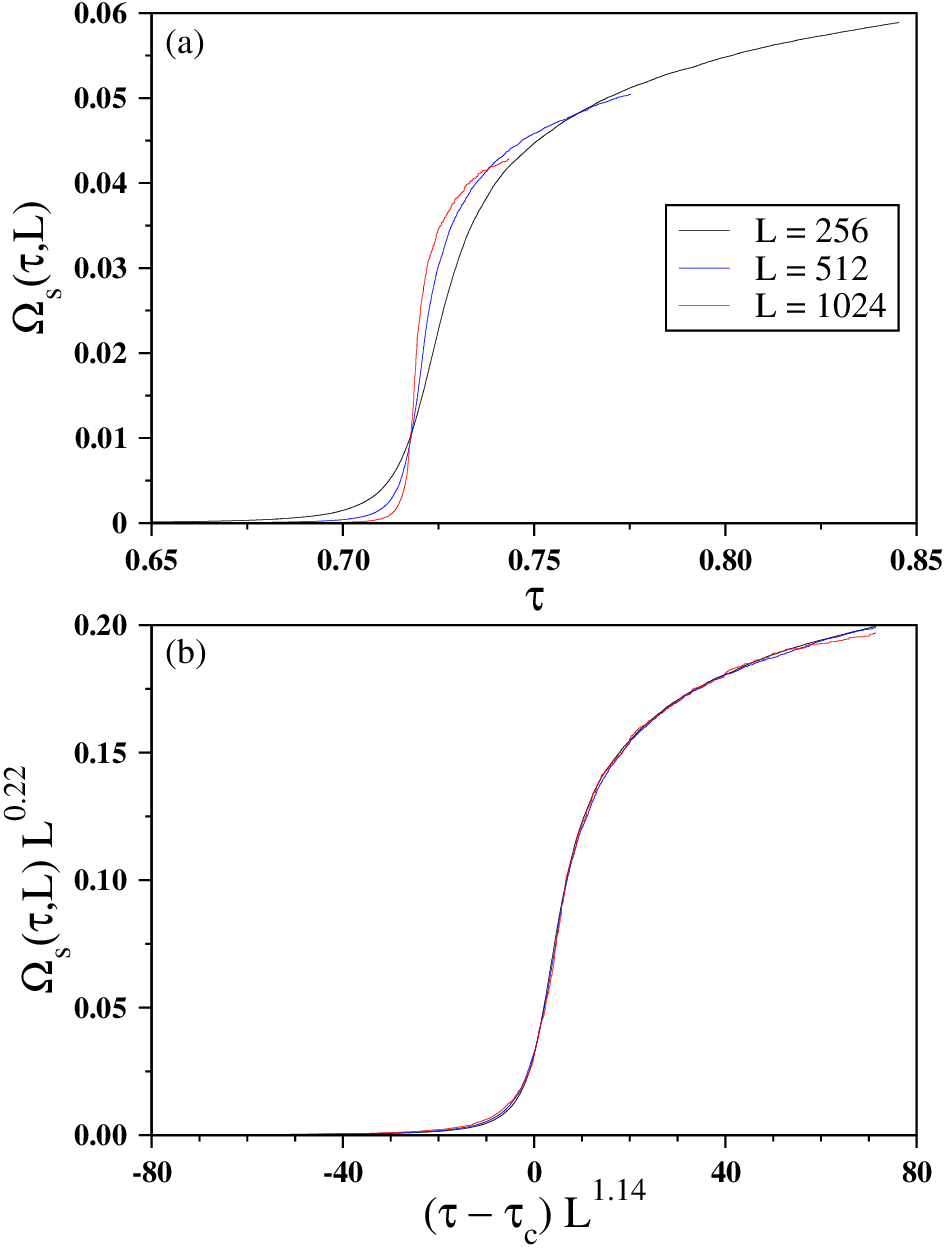}
\caption{Manna sandpile: 
	(a) The order parameter $\Omega_s(\tau,L)$ has been plotted against the drop density $\tau$
	for three lattice sizes. (b) A finite size scaling analysis of the same data have been performed where        
	$\Omega_s(\tau,L) L^{0.22}$ have been plotted against $(\tau - \tau_c)L^{1.14}$. The data collapse have
	been found to be quite good.
}
\label{FIG12}
\end{figure}

      The next question we would like to ask, what should be the standard deviation of the avalanche
   sizes for any value of $\tau$? We have simulated again a large number of independent runs for three 
   different lattice sizes. The standard deviation 
   $\sigma_s(\tau,L) = \{\langle s(\tau)^2 \rangle - \langle s(\tau) \rangle^2 \}^{1/2}$
   have been calculated for each value of $\tau$ and plotted in Fig. \ref {FIG09}(a) using different colors for different
   lattice sizes. The values of $\sigma_s$ grow steadily in the sub-critical stage but fluctuate around a steady 
   value in the stationary state. This steady value increases rapidly as the lattice size $L$ increases. We have therefore
   re-plotted the scaled values in Fig. \ref {FIG09}(b). On plotting $\sigma_s(\tau,L) / L^{2.333}$ against $(\tau - \tau_c)L$
   a nice data collapse has been observed that implies $\sigma_s(\tau) \sim (\tau - \tau_c)^{-2.333}$ in the asymptotic
   limit of large $L$.

\section {Manna sandpile}

      The dynamical evolution process in Manna sandpile is stochastic \cite {Manna}. The threshold height for stability 
   is assumed to be $h_s = 2$. Therefore, if the number of sand particles at a site $(i,j)$ is $h_s$ or more, then there 
   is a toppling. In a toppling some particles of the toppling site are transferred randomly to the neighbouring sites 
   selecting them with uniform probabilities. There are two versions of the Manna model depending on if all the particles
   or only 2 particles are transferred. The second model has been shown to have the Abelian property \cite {Dhar3}. In 
   the stationary state, the average density of particles per lattice site of the stable state is $h_c$ = 0.7170(4) \cite {Huynh}
   for the abelian Manna model.

\begin{figure}[t]
\includegraphics[width=6.0cm]{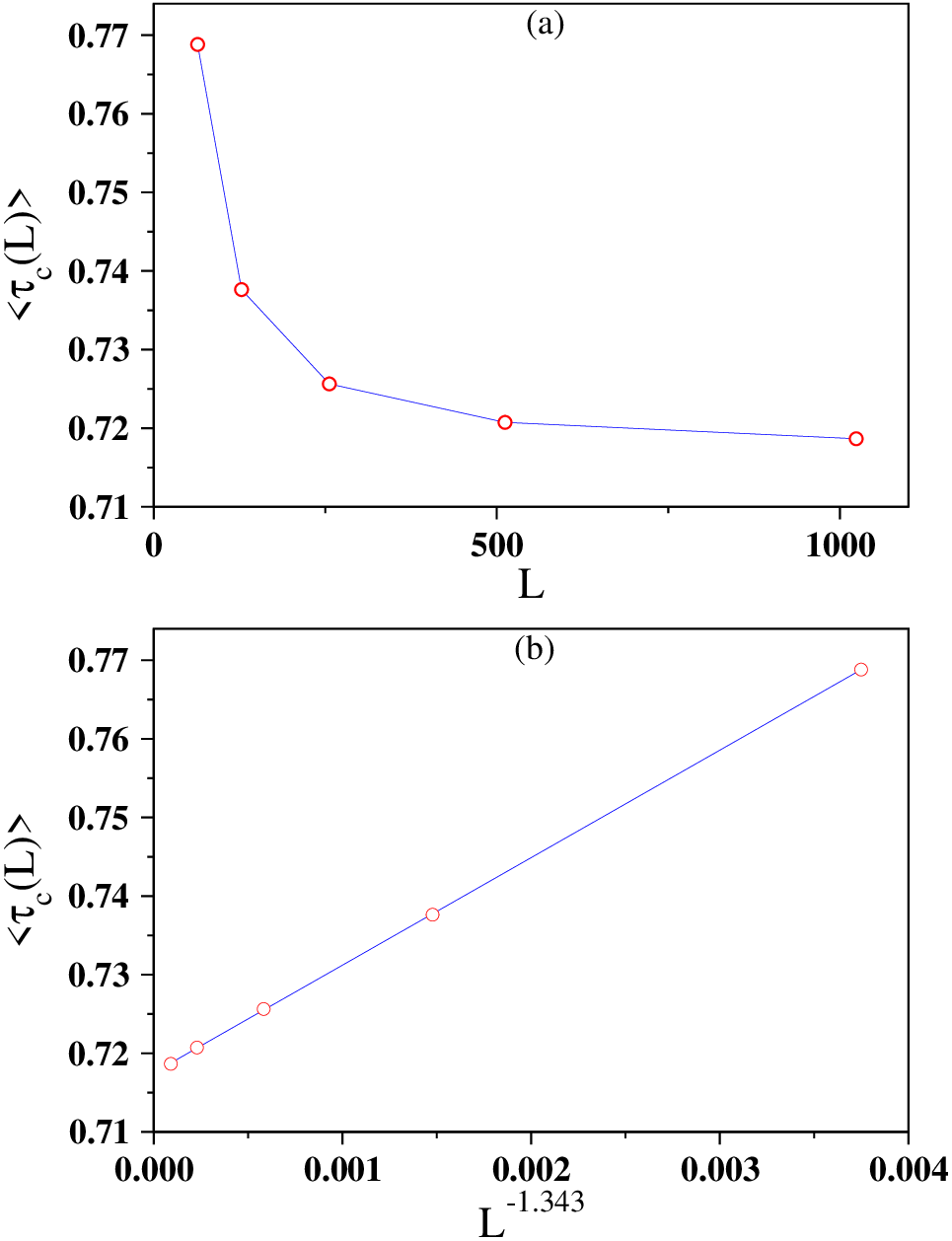}
\caption{Manna sandpile:
   (a) The average value of the critical drop density $\langle \tau_c(L) \rangle$ when the order 
       parameter $\Omega_d(\tau,L)$ gets its largest jump plotted against the lattice size $L$.
   (b) The same values of $\langle \tau_c(L) \rangle$ have been plotted against $L^{-1.343}$ to
       obtain the best fitted straight line which extrapolates to $\tau_c = 0.7175$ and $1/\nu$= 1.343.
}
\label{FIG13}
\end{figure}

      Here again we have estimated the critical drop density in two different ways. In the first case the spanning avalanches
   are considered. The average drop density $\langle \tau_c(L) \rangle$ corresponding to the first time spanning avalanches
   have been plotted in Fig. \ref {FIG10}(a) against the lattice size $L$. On increasing the system size, the critical value
   of the drop density decays. In the adjacent Fig. \ref {FIG10}(b) same values of $\langle \tau_c(L) \rangle$ have been 
   plotted against $L^{-1.140}$ to obtain the best fitted straight line: 
\begin{equation}
	\langle \tau_c(L) \rangle = \tau_c + const.L^{-1/\nu}.
\end{equation}
   On extrapolating to the $L \to \infty$ limit we get $\tau_c = 0.7170$ which is to be compared with the best estimate of 
   critical particle density $h_c = 0.7170$ quoted above. The excellent agreement shows that also for the abelian Manna model 
   the net outflow has a zero measure in the asymptotic limit of large system sizes and therefore it is not a relevant factor 
   for determining the critical point. Secondly, the inverse of the correlation length exponent is therefore $1/\nu = 1.141$.

      Two more related quantities have also been estimated at the critical point. In Fig. \ref {FIG11} we have plotted 
   the average of the first spanning avalanche size $\langle s(\tau_c,L) \rangle$ at the critical point 
   $\tau = \tau_c(L)$ against the lattice size $L$ using the double
   logarithmic scale. Five data points starting from $L$ = 64 to 1024 fit very nicely to a straight line 
   implying $\langle s(\tau_c,L) \rangle \sim L^{\zeta}$ with $\zeta = 2.745$. Secondly, the average area $\langle a(\tau_c,L) \rangle$ 
   of the spanning clusters have also been plotted in the same figure and the straight line fit here again implies a
   power law growth $\langle a(\tau_c,L) \rangle \sim L^D$ with $D = 2.047$. Since these avalanches are compact
   clusters, their fractal dimensions should be equal 2. Our estimate of a slightly larger value is due to a slight
   curvature in the small $L$ regime.

\begin{table}[t]
\centering
\caption{Comparison of values of $\tau_c$ and $1/\nu$ estimated by different methods.}
\begin{tabular}{|c|cc|cc|cc|} \toprule
	Method      & \multicolumn{2}{c|}{Spanning} &\multicolumn{2}{c|}{Max Jump}&\multicolumn{2}{c|}{Burning}\\ \hline
	Sandpile    & $\tau_c$ & $1/\nu$ & $\tau_c$ & $1/\nu$ & $\tau_c$ & $1/\nu$ \\ \hline
	BTW         & 2.1252   & 1.241   & 2.1258   & 1.335   & 2.1616   & 0.506   \\
	Manna       & 0.7170   & 1.141   & 0.7175   & 1.343   &          &         \\ \hline
\end{tabular}
\end{table}

      The order parameter $\Omega_s(\tau,L)$ for the avalanches for the abelian Manna sandpile have been plotted against
   the drop density $\tau$ in Fig. \ref {FIG12}(a) for $L$ = 256, 512 and 1024. Three curves get separated from one another 
   beyond the critical point $\tau_c(L)$. However, on suitable scaling of both the axes as $\Omega_s(\tau,L) L^{0.22}$
   against $(\tau - \tau_c)L^{1.14}$ we obtain an excellent collapse of all three curves (Fig. \ref {FIG12}(b)). This analysis 
   implies a scaling form:
\begin {equation}
\Omega_s(\tau,L) L^{0.22} \sim {\cal F}_M[(\tau - \tau_c)L^{1.14}].
\label {EQN08}
\end {equation}
   In the asymptotic limit of $L \to \infty$ the system size independent functional form should be 
   $\Omega_s(\tau) \sim (\tau - \tau_c)^{\beta}$ with -0.22 + 1.14 $\beta$ = 0 implying $\beta$ = 0.193.

      In the second method we have studied the avalanches of size $d$. As before, the order parameter 
   $\Omega_d(\tau,L) / L^2$ defined by the size of the largest avalanche has many discontinuous jumps. 
   Among them the position $\tau_c(L)$ of the largest jump may be considered as the critical drop 
   density for a specific run. An average over the large number of independent runs yield the value 
   of $\langle \tau_c(L) \rangle$. In Fig. \ref {FIG13}(a) we have plotted $\langle \tau_c(L) \rangle$ 
   for $L$ = 64 to 1024. Next, the same data have been plotted against $L^{-1.343}$ in Fig. \ref 
   {FIG13}(b) to obtain the best fitted straight line: $\langle \tau_c(L) \rangle$ = $\tau_c + 13.675 L^{-1.343}$, giving $\tau_c$ = 0.7175.

      Finally, we have calculated the probability distributions $D(s,L)$ for the avalanche sizes for avalanches
   (a) at $\tau_c = 0.7170$ and (b) the avalanche size just before the first time spanning avalanches. In both cases
   the distributions are very much well behaved in the sense they obey the finite size scaling analysis very well.
   A plot of $D(s,L)L^{\zeta}$ vs. $s / L^{\eta}$ shows an excellent data collapse with $\zeta = 3.5$ and $\eta = 2.75$
   in both cases (a) and (b) mentioned above. From these two exponents we get the avalanche size exponent $\kappa = \zeta / \eta$
   $\approx$ 1.272 which are in excellent agreement with the exponent values of the Manna model estimated before.

      The errors associated with our estimates of the critical drop densities are determined by the standard deviations
   of these quantities from the large number of independent measurements. Our estimates for the errors of the critical
   densities are of the order of 0.0002 and for the critical exponents and the finite size correction exponents are 
   of the order of 0.001. The following table summarizes and compares different critical exponents estimated in this 
   paper.

\section {Summary}

      In the phenomenon of self-organized criticality an externally driven non-equilibrium 
   system is studied that develops a current through the system. For example, in the sandpile model the 
   current of sand particles is created by adding sand grains in random locations. As a result, 
   inhomogeneity in particle density
   within the system is inevitably produced. That is why the dynamical evolution mechanism of 
   sandpiles include diffusion of particles due to toppling of sand columns whose heights are 
   too high. The main purpose of the diffusion process is to smooth out the sand landscape 
   through a self-organizing process. Extensive research in the topic of SOC have revealed 
   that long ranged spatio-temporal fluctuations characterize the steady state, and thus the 
   steady state is referred to be `critical'.

      We first got the clue that the regularly driving the system by adding sand particles one 
   at a time may quite well be looked upon as the control variable required in a continuous phase 
   transition. Consequently, the sandpile model is analyzed using the framework of percolation 
   transition, where the `drop density' $\tau$ is defined as the control variable, similar to how 
   a control variable is used in continuous phase transitions.

      The paper describes how the size of the largest avalanche, scaled by the system size, is 
   defined as the order parameter $\Omega_s(\tau,L)$. The sandpile model is then redefined to 
   exhibit a continuous transition from a disordered, uncorrelated system to an ordered, correlated 
   system. The study focuses on the Bak-Tang-Wiesenfeld (BTW) and Manna sandpiles, using finite 
   size scaling analysis to estimate critical exponents and critical drop densities $\tau_c$. The 
   conclusion mentions the presentation of a table of critical points and exponents.

      The implications of recognizing the sandpile model of self-organized criticality as a 
   continuous phase transition where the drop density i.e., the net flux of the sand mass per lattice
   site, is continuously increased is quite strong. The entire machinery of continuous phase transitions
   developed over many decades for many different systems may now be applicable to SOC as well. This
   would lead to applying the well known scaling theory and defining new critical exponents. In this 
   paper we have defined such exponents in Eqns. (3), (5), and (8). There are a lot more study to be
   done which remain to be explored. Overall, this study would lead to enhanced understanding of the
   self-organized critical phenomena.

      To summarize, the paper effectively connects concepts from SOC, and phase transitions in percolation theory, 
   demonstrating how the sandpile model of SOC can be analyzed as a continuous phase transition.

\section*{Acknowledgement}

      I am very much thankful to P. Grassberger, D. Dhar and B. K. Chakrabarti for many interesting and thoughtful comments.

\end {document}